\documentclass[twocolumn,showpacs,superscriptaddress,reprint,floatfix,aps,prl]{revtex4-1}

\usepackage{amsmath,amssymb,amsthm}
\usepackage{graphicx}
\usepackage{lineno}

\usepackage{hyperref}
\hypersetup{
    colorlinks=true,
    linkcolor=cyan,
    filecolor=magenta,      
    urlcolor=blue,
    citecolor=blue,
}

\usepackage{cleveref}
\usepackage{comment}

\usepackage{graphicx}
\usepackage{amsmath}

\usepackage{dcolumn}
\usepackage{bm}

\newcommand{\sst}[1]{\scriptscriptstyle{#1}}

\begin{document}

\preprint{APS/123-QED}

\title{Fluid mechanics of mosaic ciliated tissues}

\author{Francesco Boselli }
\email{fb448@cam.ac.uk}
\affiliation{Department of Applied Mathematics and Theoretical Physics, Centre for Mathematical 
Sciences, University of Cambridge, Cambridge CB3 0WA, United Kingdom}
\author{Jerome Jullien}%
\email{jerome.jullien@inserm.fr}
\affiliation{Wellcome Trust/Cancer Research UK Gurdon Institute, Tennis Court Road, Cambridge CB2 1QN,
United Kingdom}
\affiliation{Department of Zoology, University of Cambridge, Cambridge CB2 1QN, United Kingdom}
\affiliation{Inserm, Nantes Universit{\'e}, CHU Nantes, CRTI-UMR 1064, F-44000 Nantes, France}
\author{Eric Lauga}
\email{e.lauga@damtp.cam.ac.uk}
\affiliation{Department of Applied Mathematics and Theoretical Physics, Centre for Mathematical 
Sciences, University of Cambridge, Cambridge CB3 0WA, United Kingdom}
\author{Raymond E. Goldstein}
\email{R.E.Goldstein@damtp.cam.ac.uk}
\affiliation{Department of Applied Mathematics and Theoretical Physics, Centre for Mathematical 
Sciences, University of Cambridge, Cambridge CB3 0WA, United Kingdom}

\date{\today}

\begin{abstract}
In tissues as diverse as amphibian skin and the human airway, the cilia that propel fluid 
are grouped in sparsely distributed multiciliated cells (MCCs).  We investigate 
fluid transport in this ``mosaic" architecture, with emphasis on the 
trade-offs that may have been responsible for its evolutionary selection.
Live imaging of MCCs in embryos of the frog {\it Xenopus laevis} shows 
that cilia bundles behave as active vortices that
produce a flow field accurately represented by a local force applied 
to the fluid.  
A coarse-grained model that self-consistently couples bundles to the ambient flow 
reveals that hydrodynamic interactions between MCCs limit their rate of work so that 
when the system size is large
compared to a single MCC, they best shear the tissue at low area coverage,
a result that mirrors findings for other sparse distributions such as cell 
receptors and leaf stomata. 
  
 \end{abstract}

\maketitle
An indication of the importance of fluid mechanics in biology is the 
remarkable degree to which the structure of eukaryotic cilia has been conserved over the 
past billion years \cite{cilia_conserved,Ainsworth}.  These hairlike appendages 
provide
motility to microorganisms \cite{Mitchell,GoldsteinARFM} 
but also direct fluid flow inside animals during development \cite{LRO, Gallaire2020, Ferreira2017} and in mature
physiology in areas from the reproductive system \cite{Fauci} to the brain 
\citep{faubel2016}.  The two extremes of this organisimal spectrum have a fundamental distinction. In 
unicellulars like {\it Paramecium}, cilia are uniformly and closely spaced 
on the cell surface \cite{Paramecium}, while in animals they are often 
grouped together in dense bundles on multiciliated cells (MCCs) \cite{Brooks2014}
that are sparsely distributed on large epithelia, as in the trachea and 
kidney \citep{Liu2007,Vasilyev2009}.
This difference reflects the need in animal tissues to share surface area with cell types 
having other roles, such as mucus secretion.

\begin{figure}[b]
    \centering
   \includegraphics[width=\linewidth]{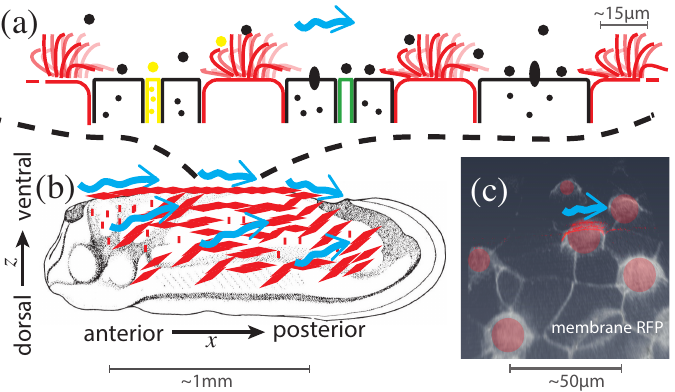}
    \caption{Ectoderm of embryonic \textit{Xenopus laevis} at tailbud stages. (a) Schematic side view of MCCs 
    (red) intermixed with secreting cells. (b) Location of 
        MCCs across the embryo (adapted from \cite{NFstages,Xenbase}) and cilia-driven flow (blue arrows). 
        (c) Confocal image of cell membranes (stained by membrane-RFP), with MCCs segmented in red, in ventral 
        region of skin.}
    \label{fig:Embryo}
\end{figure}

The workings of cilia bundles and the significance of their sparse ``mosaic" pattern 
for fluid transport have only begun to be investigated, primarily limited to {\it in vitro} or 
{\it ex vivo} studies \cite{Viallat1,Viallat2,Prakash}.
Here we address the fluid mechanics of mosaic tissues using 
embryos of the amphibian \textit{Xenopus laevis} in which, by analogy to human airways, 
cilia driven flow sweeps away mucus and trapped pathogens (Fig.\ \ref{fig:Embryo}).
To date, the flow has served 
as a readout of cilia beating in the study
of tissue patterning and cilia disorders \cite{Twitty1928,Deblandre1999,Werner:2013}; here we take advantage of
the geometry of {\it Xenopus} embryos to obtain side views
of cilia bundles and quantify the flows they drive.  
As those cilia collectively sweep
through cycles consisting of an extended ``power" stroke and compact 
``recovery" stroke  close to the surface \cite{cilia_strokes},
the flow within each bundle appears as an active vortex. While the flow driven by 
a single such vortex 
decays quickly with distance from the skin, a coarse-grained model shows that 
long range contributions of other bundles slows the decay of this {\it endogenous} flow 
and determines the shear stress at non ciliated cells. 
From measurements of beating changes induced 
by {\it exogeneous} flows, we determine linear response coefficients describing the 
coupling between forces applied by bundles and the flows they generate; we find
that hydrodynamic interactions between MCCs lead to maximization at low area coverage 
of shear at the intervening tissue.  These results thereby suggest an explanation for
the low area coverages observed in nature.  

\begin{figure*}[t]
\centering
\includegraphics[width=1.0\linewidth]{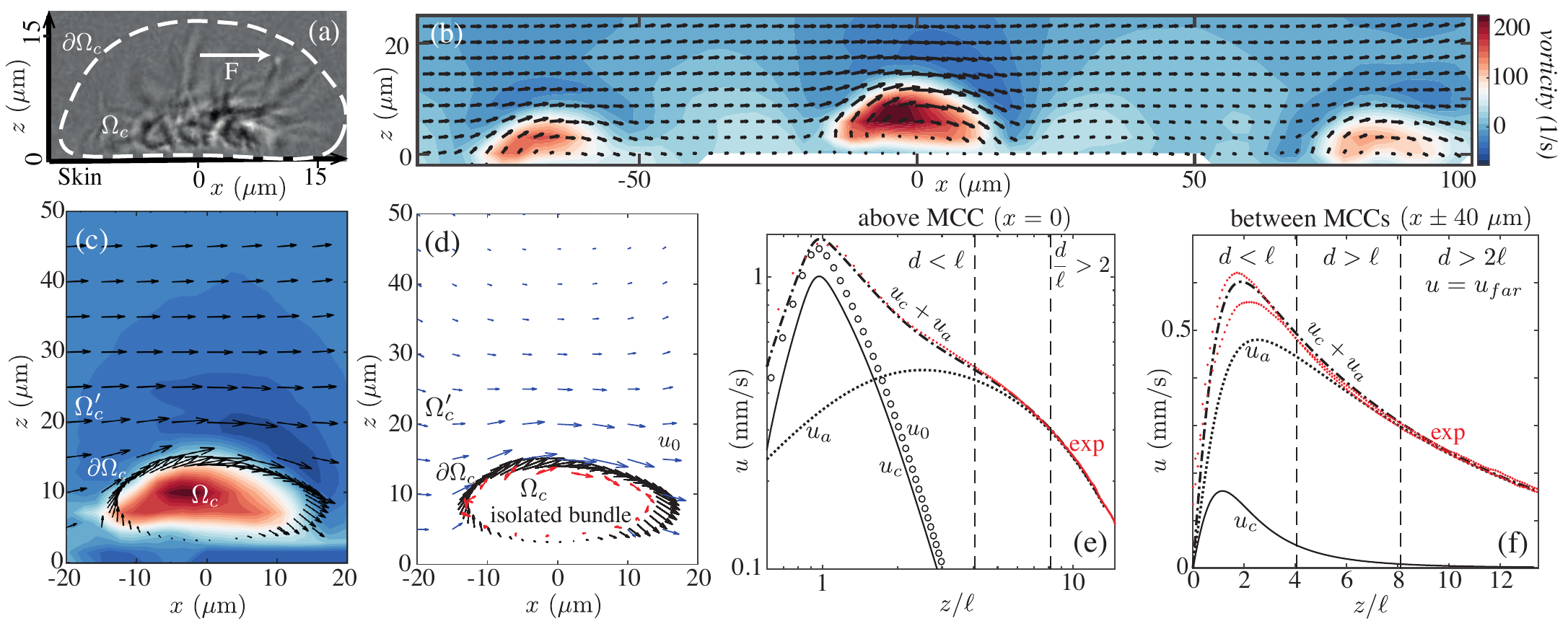}
\caption{Flow field around multiciliated cells. (a) Lateral view of an MCC showing (dashed)
path of cilia tips and applied force $\mathbf{F}$.  
(b) Experimental velocity field and vorticity in a plane normal 
to skin near several MCCs, with $z$ into fluid and $x$ along flow. (c) Near an MCC, as in (b), with direction of cilia 
tip motion (black arrows) on $\partial \Omega_c$.
(d) Estimated flow field $u_0$ for an isolated MCC (blue arrows):   
Stokeslets (red arrows) are used to fit velocity near cilia tips. Lateral velocity as a function 
of $z$ at (e) $x,y=(0,0)$ and 
(f) $(\pm40\,\mu{\rm m},0)$ 
in experiment (exp) and theory, with $u_0$ driven by an isolated bundle and $u_c$ by a bundle 
exposed to the endogenous flow $u_a$ (see also Figs. S1-S3 \cite{SM}).}
\label{fig:flowfield}
\end{figure*}

The epidermis of \textit{Xenopus} 
has strong similarities with human mucociliary epithelia.
The sparsely located MCCs whence emanate hundreds of cilia that 
drive a homogeneous anterior-to-posterior, A-P or head-to-tail, flow (Fig.~\ref{fig:Embryo}) are 
surrounded by non-ciliated cells secreting mucus-like material \cite{Nagata2005}, 
including ``goblet cells" that cover most of the tissue, mosaically scattered small 
cells \cite{Dubaissi2011,Dubaissi2014} secreting serotonin vesicles that modulate the ciliary 
beat frequency \cite{Walentek2014}, and ionocytes transporting ions important for homeostasis. 

Wild-type \textit{Xenopus laevis} embryos were obtained via {\it in vitro} fertilization 
\cite{hormanseder2017,SM}, and grown in $0.1\times$ Modified Barth's Saline 
at room temperature (or 15$^\circ$ C to reduce the growth rate, if required).  They were imaged at 
stage $28$ \cite{NFstages} after treatment with a minimal dose of anaesthetic ($\sim 0.01\%$ Tricaine) to avoid twitching 
(without affecting cilia dynamics \cite{Werner:2013}).  
Embryos lie on one of their flat flanks at this stage, providing a side 
view of cilia bundles of ventral MCCs (Fig.~\ref{fig:Embryo}), whose power strokes are in the A-P direction 
(left to right in figures) so cilia and the flows stay mostly within the focal plane. 

In flow chamber experiments, embryos were perfused with a peristaltic pump while in a 
Warner Instruments chamber (RC-31A): a $4$ mm $\times$ $37$ mm channel cut into a $350\,\mu$m 
thick silicon 
gasket sandwiched between two coverslips that keep the embryo in 
place by pressing against its sides. 
The dorsal part of the embryo was positioned closer to the chamber wall, the anterior region of 
interest was $>\! 2\,$mm away, and the A-P axis parallel to the channel axis, 
the main direction of the perfusing flow.  Brightfield images of cilia and $0.2-0.5\, \mu$m tracers 
(mass fraction $\sim0.01 \%$) were acquired 
at $2000$ frames/s for $\ge 1\,$s by a high speed camera (Photron Fastcam SA3) on an inverted microscope 
(Zeiss Axio Observer) with a long distance $63\times$ objective (Zeiss LD C-Apochromat). 
Images were filtered by subtracting their moving average.
Flow fields $\mathbf{u}=(u,v,w)$ were estimated by Particle Image Velocimetry 
(PIVlab) and averaged over time.

We set the stage by summarizing the important length and time scales.
MCCs are spaced apart by $40-80\,\mu$m and uniformly distributed with average density 
$\mathcal{P} \approx 2.6\times 10^{-4}\,\mu$m$^{-2}$, giving an average spacing $d=\sqrt{1/\mathcal{P}}\sim 
62\, \mu$m. With $\ell \sim 15\, \mu$m ($\bar{\ell}=14.52 \pm 0.21\mu$m) the cilia length, 
the average cellular area 
$\sim 287 \pm 11 \, \mu$m$^2$
is $\sim \ell^2$, which gives a coverage fraction 
$\phi = (\ell/d)^2 \sim 0.07$.  
The cilia on MCCs beat at a frequency $f\sim 20-30\,$Hz and during a power stroke their
tips move a distance $\sim 2\ell$ [Fig.~\ref{fig:flowfield}(a)] in half a period, reaching speeds 
$V_c\sim 4f\ell \sim 1\,$mm/s,  so the 
Reynolds number $\rho V_c\ell/\mu$ (with $\rho$ the density and $\mu$ the viscosity of water) is
$\sim\! 0.01$, well in the Stokesian regime.  
Using the fluid speed $u_v\sim 0.5\,$mm/s between vortices as typical of the periciliary region, the 
P{\'e}clet number $u_v\ell/D > 1$ even for small molecules.

Cilia within an MCC are not synchronized; their tips move in a tank-treading manner \cite{SM}, 
generating vorticity 
${\bm{\omega}}\parallel \mathbf{e}_y$ perpendicular to the beating plane.  Each MCC is thus an 
{\it active vortex}, as seen in Figs.~\ref{fig:flowfield}(b,c).  The vorticity can exceed
$\sim 150\,$s$^{-1}\sim 2V_c/\ell$, is colocalized with the cilia, and rapidly diffuses 
at larger $z$ as the flow becomes parallel to the skin. Above non-ciliated cells between MCCs, 
there is a shear flow for $z<\ell$, while further away ($z \gtrsim d$), the discreteness of the MCCs 
is washed out by viscosity and the horizontal velocity $u$ is independent of $x$ 
and falls off slowly with $z$ [Figs.\ \ref{fig:flowfield}(e,f)].

The first step toward understanding the coupling between cilia beating and fluid flow involves
quantifying the contribution of a single MCC. We introduce a boundary 
$\partial\Omega_c$ enclosing the volume $\Omega_c$ of the active vortex [Fig.~\ref{fig:flowfield}(a)], and
extrude it in $y\in [-10\, \mu{\rm m}, 10\, \mu{\rm m}]$, the measured size of the vortex. 
The Stokes equations in the complement $\Omega_c'$, are solved using an envelope 
approach \cite{Blake1971a,Brumley2016} in which the dynamics of the cilia tips determine the flow, noting that 
the flow from an isolated cilium is well-approximated in the far field as that of a 
point force (Stokeslet) \cite{Brumley2014}.  We position $N$ Stokeslets at $\mathbf{s}_n=(x_n,y_n,h_n)$ 
in $\Omega_c$ and find their strengths $\mathbf{f}_n=(f_{n,x},0,f_{n,z})$ by fitting the 
velocity on $\partial \Omega_c$ and a no-slip boundary at $z=0$ [Fig.~\ref{fig:flowfield}(d)]. 
This gives an estimate of the flow  $\mathbf{u}_0$ driven by a bundle in an
otherwise quiescent fluid. 
The $x$-components are of interest, as they alone contribute to net AP flow. For $z>2l$, the 
component $u_0$ falls off for $z\gg h$ like the flow $u_s(z)\sim 3Fh^2/4\pi\mu z^3$ above a single 
Stokeslet parallel to 
and a distance $h$ above a no-slip wall \cite{Blake1971} (Fig.\ S1 \cite{SM}).
The data in Figs.~\ref{fig:flowfield}(e,f) show a much weaker fall-off than $z^{-3}$ for 
$z>2l$.
This slow decay arises from the long range contribution of more distant MCCs, as we now show. 

The flow $u_s$ due to a force $f_x$ can be approximated by its 
far field limit. In cylindrical coordinates ($\rho,z$) centered at the bundle, 
$u_s \approx hf_x\tilde{S}_{11}(\rho,z)$,  with
\begin{equation}
	\tilde{S}_{11}(\rho, \theta, z)= \frac{3 }{2\pi\mu}\frac{z\rho^2 \cos^2{\theta}}{(\rho^2+z^2)^{5/2}}.
    \label{single_stokeslet}
\end{equation}
The Stokeslet contributions can be lumped into an effective force 
    $F_c= \ell^{-1}\sum_n{h_n f_{n,x}}$, which, applied at $z=\ell$, matches the far field 
    $u_s \approx \ell F_c \tilde{S}_{11}$.
This also applies for an effective  local moment 
(rotlet) $\Gamma_c =2\ell F_c$ \cite{Blake1974}.

The observed velocity $u(z)\mathbf{e}_x$ in the region $z>d$ is independent of 
$x$ (Fig. S2 \cite{SM}) and can be described most simply in a model of the 
skin as a uniform 
distribution of $x$-directed Stokeslets with area density ${\cal P}$. 
Summing the contributions of all bundles up to a cutoff radius $\Lambda$ that incorporates finite 
embryo size, we obtain
\begin{equation}
    u_{\rm far}(z;\Lambda)={\cal P}F_c\int_0^{2\pi}\!\int_0^\Lambda \! \rho\, d\rho\, d\theta\,  \tilde{S}_{11}(\rho,\theta,z),
    \label{ufar}
\end{equation}
which has the scaling 
form $u_{\rm far}=V G(z/\Lambda)$, with 
\begin{equation}
\label{eq:VL}
G(\chi)= 1-\frac{3\chi+2\chi^3}{2(1+\chi^2)^{3/2}}
\end{equation} 
and $V={\cal P}F_c l/\mu$.  
$G$ decreases monotonically from $G(0)=1$ to $G(\infty)=0$. 
For any fixed $z$, as the organism size $\Lambda\to \infty$, $\chi\to 0$, giving a flow
independent of $z$ with speed $V$ \cite{Osterman2011}, while for any fixed lateral scale $\Lambda$, the
asymptotic flow field vanishes as $z/\Lambda\to \infty$. For the fitted parameters $V\simeq 0.64\,$mm/s 
and $\Lambda=300\, \mu$m,  $u_{far}$ provides an almost perfect fit of the data in 
Figs.\ \ref{fig:flowfield}(e,f) for  $z>2d$ (also Fig. S2 \cite{SM}). Direct
summation of discrete Stokeslets on a lattice produces nearly identical 
results, validating the approximation of a continuous distribution for the far-field flow
(Fig.\ S3 \cite{SM}). For this value of 
$V$ and the observed density ${\cal P}$, we have the far-field estimate $F_c\simeq 159\,$pN. 
A slightly larger effective force $F_0\simeq 200\,$pN is obtained by solely fitting 
the velocity at the cilia tips (Fig.\ \ref{fig:flowfield}d), which is to be expected as
this approach does not account for the endogenous flow $u_a \mathbf{e}_x$ to which 
a bundle is self-consistently exposed.  

We estimate the endogeneous ambient flow $u_a$ by subtracting from $u_{\rm far}$ the contribution 
from a single cilia bundle, taken as a 
distribution of radius $d/2$, 
\begin{equation}
    u_a \approx V[G(z/\Lambda)-G(2z/d)].
    \end{equation}
As $u_c = u - u_{a}$ represents the contribution of a single bundle, we 
test the far-field estimate with a near-field fit of $u_c$ in $\Omega_c'$ by means of 
Stokeslets within  $\Omega_c$. Doing so for the volume ($-15<x<15,-10<y<10, 0<z<d$),  
the superposition  $u_c + u_{a}$ gives an excellent fit to the data in Figs. \ref{fig:flowfield}(e,f),
above and between bundles, with 
$F_c \simeq 160\,$pN, nearly identical to the far-field estimate.  

\begin{figure*}[t]
\includegraphics[width=1.0\linewidth]{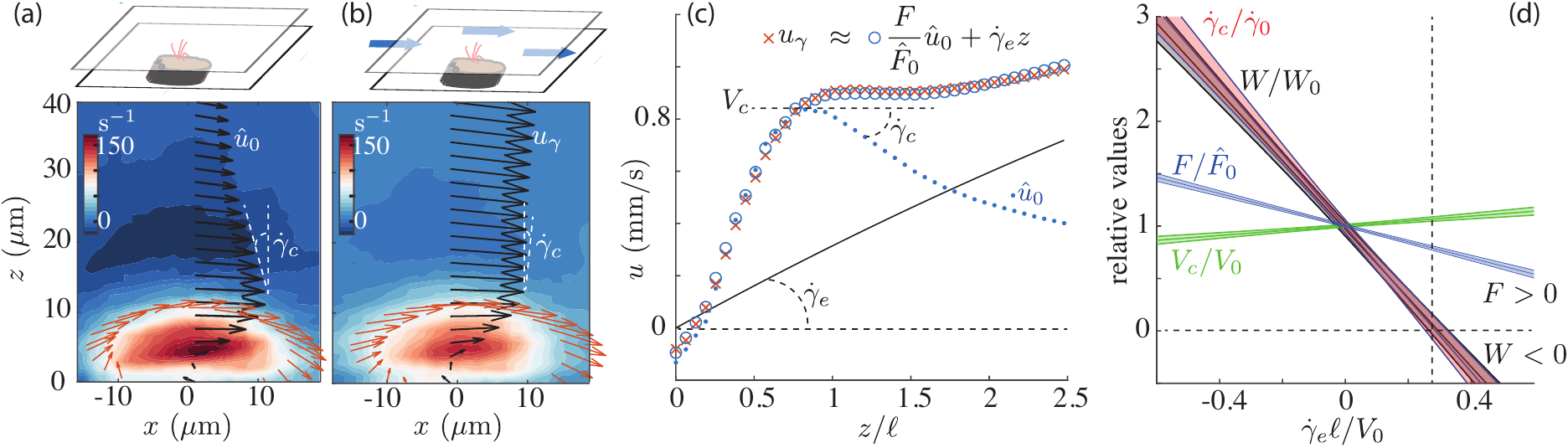}
\caption{Response of a cilia bundle to shear flow.  
(a,b) Vorticity contours and velocity vectors before and during perfusion. (c) Velocities $\hat{u}_0$ and $u_\gamma$, 
exogenous shear flow $\dot{\gamma}_e z$, and linear combination $(F/\hat{F}_0)\, \hat{u}_0 + \dot{\gamma}_e z$ fitting $u_\gamma$.  
(d) Linear fits of the variation of estimated force $F$, velocity $V_c$ and shear rate 
$\dot{\gamma}_c =\partial u/\partial z$
measured above cilia tips, and rate of work $W\propto \dot{\gamma}_c V_c$ (overlaping $\dot{\gamma}_c$), 
normalized by values at $\dot{\gamma}_e=0$.  Shaded regions are $95\%$ confidence intervals of averages
over $10$ samples.}
\label{fig:FlowChamber}
\end{figure*}

For comparison, the average 
lateral force generation over one cycle of beating (assuming only the power stroke contributes)
can be estimated from resistive force theory \cite{RFT} as 
$f\sim \zeta_{\sst\perp}\ell V_c/12\simeq 3\,$pN \cite{SM}, 
where $\zeta_{\sst\perp}=4\pi\mu/\vert\ln(\sqrt{e}\varepsilon)\vert$ 
is the transverse drag coefficient for a slender filament of aspect ratio $\varepsilon$ 
(for cilia, $\varepsilon\sim 75$).
We infer from the estimated $F_c$ that the effective number of cilia contributing to the 
Stokeslet is $\sim 53$, about half the typically $\sim\!\! 100$ cilia in an MCC, 
reflecting force cancellations from phase shifts between cilia.

The fact that $F_c/F_0<1$ shows that it is necessary to incorporate the response of a 
cilia bundle to an ambient flow.
We probed this experimentally by exposing the bundle to an exogenous shear flow 
$\dot{\gamma}_e z \mathbf{e}_x$ in a flow chamber (Fig.\ \ref{fig:FlowChamber}).
When $\dot{\gamma}_e = 0$, the  cilia tips move with velocity $V_0$ and drag the 
fluid, generating a negative shear rate $\dot{\gamma}_0\approx -23\,$ s$^{-1}$.   Pumping 
fluid in the same direction, the shear 
rate at the tips $\dot{\gamma}_c$ decreases linearly with the hydrodynamic load $\dot{\gamma}_e$. 
The 
corresponding velocity $V_c$ tends to increase, but at a much slower rate. 
The rate of work above the cilia tip envelope $ \propto -\dot{\gamma}_c V_c$ thus decreases 
almost at the same rate as 
$\dot{\gamma}_c$, and for $\dot{\gamma}_e \ell/V_0 > 0.3$ becomes negative, consistent 
with a dissipative bundle. 

The lateral velocity $u_\gamma \mathbf{e}_x $ just above the bundle ($z<2.5 \ell$) is 
well fitted by the linear combination $u_\gamma \approx C \hat{u}_0  + \gamma_e z$ (Fig. \ref{fig:FlowChamber}c) 
with  $\hat{u}_0(\hat{F}_0)$  the profile at $\dot{\gamma}_e = 0$. Note that  $C\approx F/\hat{F}_0$, 
as confirmed for the above calculations where  $u_c \approx (F_c/F_0) u_0$ for $z>\ell$ 
(Fig.\ S2(b) \cite{SM}). 
The slope of $F/\hat{F}_0$ versus $\dot{\gamma}_e \ell/V_0$ in Fig.\ \ref{fig:FlowChamber}, 
would be unity if the bundle dynamics were fully preserved ($V_c = V_0$), and zero if the 
bundle's force were constant. The measured slope $0.76\pm 0.06$
confirms the resistive behavior and allows us to parameterize the coupling of the cilia
to the ambient flow by the linear relation
$F \approx \hat{F}_0 - \alpha \ell \dot{\gamma}_e$, with $\alpha = 0.76 \hat{F}_0/V_0$.

To close the loop on a self-consistent coupling of the bundles to the flow, we now 
replace $\dot{\gamma}_e$ with the endogenous shear $\dot{\gamma}_a = \partial{u_a}/\partial{z}|_{z=0}$,
giving $F \approx F_0 - \alpha \ell \dot{\gamma}_a$, with $F_0$ again the effective force applied 
by a bundle in 
an otherwise quiescent fluid. Since $\dot{\gamma}_a \approx 3 V/d [1 - {\cal O}(d/2\Lambda)]$
for large tissues, we have $\dot{\gamma}_a\approx  3F\ell/\mu d^3 $, and thus
\begin{equation}
    F(d)=\frac{F_0}{1 + \lambda \left(\ell/d\right)^3},
    \label{eq:FoF}
\end{equation}
where the $\lambda=3\alpha/\ell\mu$ is the key parameter of
the self-consistent theory.  Using typical values ($\hat{F}_0=160\,$pN, $V_0=1.3\,$mm/s, 
$\ell=15\,\mu$m), we find $\lambda\approx 18.6$.

\begin{figure}[b]
    \centering
    \includegraphics[width =1.0\linewidth]{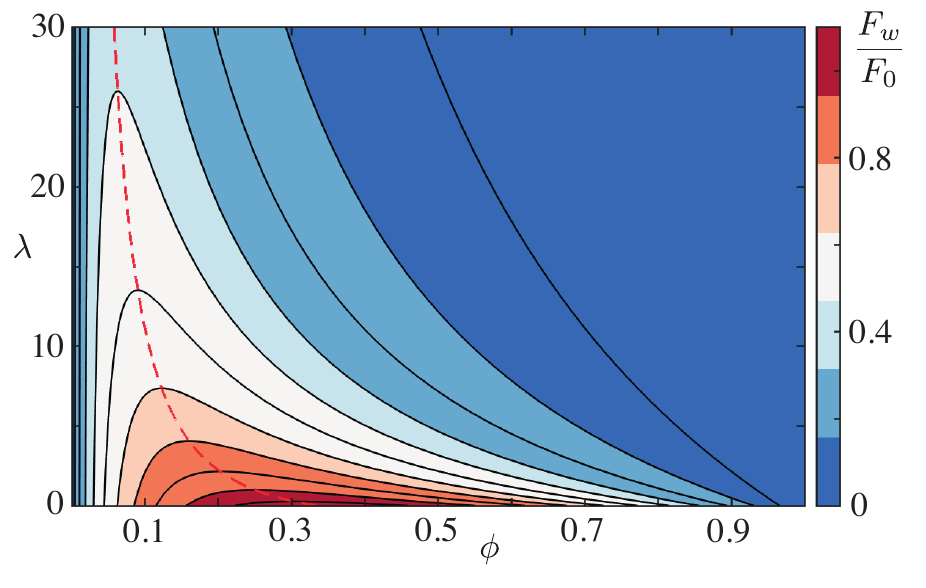}
    \caption{Force on non-ciliated cells in the self-consistent model. Contour plot 
    of \eqref{wallforce} in parameter space.  Dashed line traces optimization ridge.}
    \label{fig:four}
\end{figure}

The relation \eqref{eq:FoF} can be used to address several aspects
of sparse cilia distributions \cite{SM}. The force applied to the wall per bundle is
$\sim \mu \dot{\gamma}_a d^2=3F_0 a/(1+\lambda a^3)$, with $a=\ell/d$, and has a maximum at
$d_{max} = (2 \lambda)^{1/3} \ell\approx 50\, \mu$m as does the contribution $FV$ of a 
single bundle to the rate of work.  The force $F_w\approx \mu \dot{\gamma}_a (d^2-\ell^2)$ applied 
to the {\it non-ciliated} cells is maximal
for $d \approx 54 \,\mu$m.  Both values are in excellent agreement with those observed.
Using the relation $\phi=(\ell/d)^2$ we can express the wall force as a function of
the coverage fraction $\phi$, 
\begin{equation}
    \frac{F_w}{F_0}=\frac{3\phi^{1/2}(1-\phi)}{1+\lambda\phi^{3/2}}.
    \label{wallforce}
\end{equation}
The contour plot of $F_w/F_0$ in the $\phi-\lambda$ parameter space in Fig. 
\ref{fig:four} shows that the optimum area fraction is a strongly decreasing function
of $\lambda$ and can reach values far below unity for $\lambda\sim 20$, as in the present
study.
Extra endogenous loads $\dot{\gamma}_e<0$, as expected for internal tissues, appear  
as an additional contribution to the ambient flow $\dot{\gamma}_a+\dot{\gamma}_e$. 
These loads will contribute to \eqref{eq:FoF} 
as lower values of $\lambda$, and indeed, 
consistently larger, yet still low coverage fractions of the airways of several 
animals have been estimated to be 
$0.4-0.5$ \cite{Prakash}, qualitatively consistent with Fig. \ref{fig:four}. 
We infer that the observed mosaic patterns are close to optimal in terms of 
the clearing force applied to non-ciliated cells.  

We close with comments on connections to other systems 
with sparse distributions of active elements. 
The force applied to the outer fluid by the cilia tips on the envelope $\partial\Omega_c$ is equal 
and opposite to that applied to the skin, and we can simplify our results on the shearing of 
non-ciliated cells by reconsidering \eqref{ufar}  
as the flow of a patch of activity with given slip velocity $V$ of radius $\Lambda$. The shear stress 
driving the flow is  
$\tau_{\Lambda} \approx 3\mu V/2\Lambda$, which we assume constant over a bundle.
Setting $\Lambda=\ell$ and integrating over a tissue with $N$ bundles we obtain $J \sim N\ell^2\tau_{\ell}$.  
By contrast, if we set 
$\Lambda=R$, the local shear stress is
$\tau_R=3\mu V/2R$ and the force over the entire surface is $J_R\sim \pi R^2 \tau_R$.  With $N=\pi R^2\phi/\ell^2$, the ratio
\begin{equation}
    \frac{J}{J_R} \sim  \frac{R}{\ell} \phi
    \label{surprise}
\end{equation} 
measures how well a distribution of non-interacting MCCs shears the surface 
relative to the collection.  The linear scaling of \eqref{surprise} with $\phi$ is expected, but the large 
prefactor $R/\ell\sim 20$ 
(system size/MCC size) implies that $J/J_R$ can approach unity 
for area fractions as low as $\phi\sim \ell/R\sim 5 \% $. 
The form of this result mirrors one found by Jeffreys \cite{Jeffreys1918} for the 
evaporation rate from sparsely distributed leaf stomata, rediscovered years 
later \cite{Berg1977} in the context of ligand binding to sparse cell receptors \cite{GoldsteinPT}.

The results presented here suggest that long range hydrodynamic interactions between 
multiciliated cells allow efficient peri-ciliary transport at relatively low coverage, 
favoring the coexistence of multiple cell types in large tissues. 
This is likely just one aspect of more general mechanisms that maintain efficient transport in the 
upscaling events marking the evolutionary transition from 
unicellular to larger multicellular systems. 

\begin{acknowledgments}
This work was supported in part by Wellcome Trust Grant 101050/Z/13/Z) and 
Medical Research Council grant MR/P00479/1 (JJ), ERC Consolidator grant 682754 (EL), 
Wellcome Trust Investigator Award 207510/Z/17/Z, 
Established Career Fellowship EP/M017982/1 from the
Engineering and Physical Sciences Research Council, and the Schlumberger Chair Fund (REG). 

\end{acknowledgments}

\vfil
\eject

\section{Supplemental Material}
This file contains additional experimental and calculational details.

\setcounter{equation}{0}
\setcounter{figure}{0}
\setcounter{table}{0}
\setcounter{page}{1}
\makeatletter
\renewcommand{\theequation}{S\arabic{equation}}
\renewcommand{\thefigure}{S\arabic{figure}}
\renewcommand{\bibnumfmt}[1]{[S#1]}
\renewcommand{\citenumfont}[1]{S#1}

{\bf Embryo Culture:}
{\it Xenopus} embryos were prepared as described previously \cite{Hormanseder2017}. Briefly, mature {\it Xenopus laevis} 
males and females were obtained from Nasco \cite{Nasco}. Females were injected with $50$ units of pregnant mare 
serum gonadotropin $3$ days in advance and $500$ units human chorionic gonadotropin $1$ day in advance in the dorsal 
lymph sack to induce natural ovulation. Eggs were laid in a $1\times$ MMR buffer ($5\,$mM HEPES pH $7.8$, 
$100\,$mM NaCl, $2\,$mM KCl, 
$1\,$mM MgSO$_4$, $2\,$mM CaCl$_2$, $0.1\,$mM EDTA). {\it Xenopus} embryos were cultured at room temperature or 
$15^\circ$ C in the $0.1\times$ MMR until they reached stage $27/28$. Experiments with embryos were performed at the 
late tailbud stages (stages 28-30, as describe in Faber and Nieuwkoop \cite{NF}). Embryos were terminated humanely 
immediately following the experiments.

Our work with {\it Xenopus laevis} is covered under the Home Office Project License PPL 70/8591 and frog 
husbandry and all experiments were performed according to the relevant regulatory standard.
All experimental procedures involving animals were carried out in accordance with the UK Animals (Scientific Procedures) 
Act 1986.  Moreover, we only used surplus embryos for this study, to conform with the NC3Rs guidance to exploit 
the possibility to minimise the use of animals by sharing embryos with collaborators.

\begin{figure*}
\begin{center}
\includegraphics[width=0.6\linewidth]{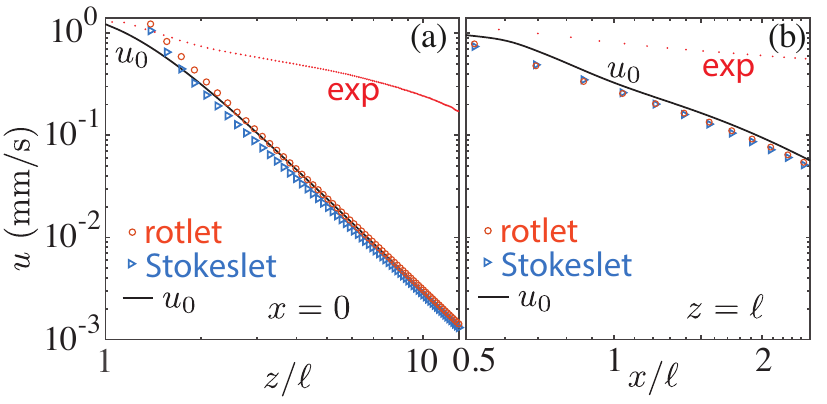}
\caption{The contribution of a single bundle of cilia decays as a single effective force $F_c$ or moment 
$2\ell F_c$, while the measured flow profile decays much more slowly due to the contributions from other MCCs 
(exp). The lateral velocity $u_0$, with reference to Fig. 2(d), is shown  (a) above the bundle as a function of 
$z$, and (b) between bundles at $z=\ell$, as a function of $x$.}
\label{fig:Sfig1}
\end{center}
\end{figure*}

\begin{figure*}[t]
\begin{center}
\includegraphics[width=0.6\linewidth]{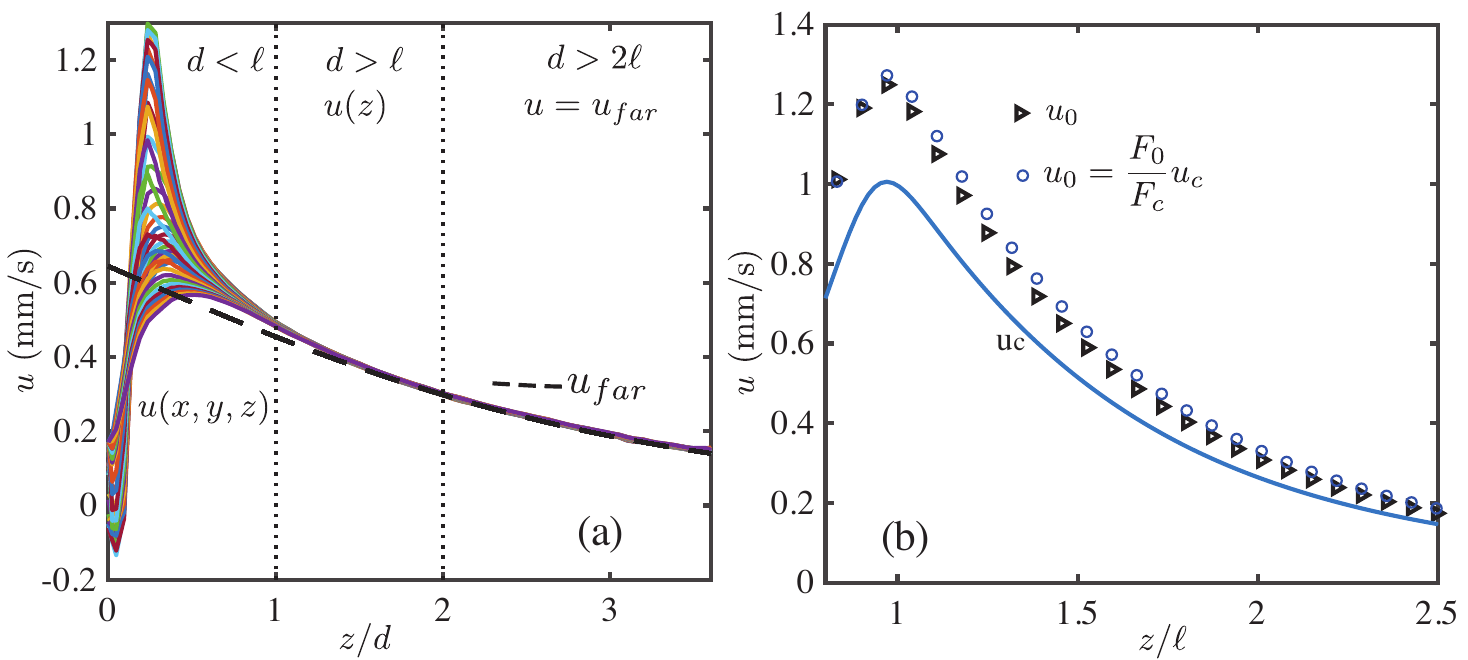}
\caption{Supplement of Fig.\ 2. (a) Measured lateral velocity $u(z)$ above and between bundles. For $z>d$, the $u(z)$ becomes independent on $x$. For $d>2d$, the $u(z)$ matches the far-field model $u_{far}(z)$ of a uniform distribution of cilia. Different colours correspond to velocity profiles at several $|x|<35$ $\mu$m. (b) Linear dependance of the near-field $u_c$ on the effective force $F_c$, for $z>\ell$.}
\label{fig:Sfig2}
\end{center}
\end{figure*}

\medskip

{\bf Statistics:}
To fit any quantity $y$ measured at a given hydrodynamic load $x$, we assume a linear relation 
$y = a + bx$ and find  $95\%$ confidence intervals for the averages $\bar{a}$ and $\bar{b}$ of the 
parameters  $a_{i}$ and $b_{i}$ given by a least-square fit of the measured values $(x_{i,m},y_{i,m})$ 
acquired for the $i$-th  MCC. We have \cite{Bevington1993}:
\begin{subequations}
\begin{align}
\bar{a}&=\frac{\sum_{i=1}^N a_i/\sigma_{a,i}^2}{\sum_{i=1}^N (1/\sigma_{a,i}^2)}\pm 2\sigma_{\mu \bar{a}}\\
\bar{b}&=\frac{\sum_{i=1}^N b_i/\sigma_{b,i}^2}{\sum_{i=1}^N (1/\sigma_{b,i}^2)}\pm 2\sigma_{\mu \bar{b}},
\end{align}
\end{subequations}
The standard errors for the mean parameters are the square roots of
\begin{subequations}
\begin{align}
\label{eq:AVGerror-a}
\sigma_{\mu \bar{a}}^2&= \frac{1}{\sum_{i=1}^N 1/\sigma_{a,i}^2} \\
\sigma_{\mu \bar{b}}^2&= \frac{1}{\sum_{i=1}^N 1/\sigma_{b,i}^2}, 
\end{align}
\end{subequations} 
while the parameter variances for the $i$-th MCC are
\begin{subequations}
\begin{align}
\sigma_{a,i}^2 &=  \frac{\sigma_i^2}{\triangle_i^{'}}\sum_{m=1}^M x_{i,m}^2\\
\sigma_{b,i}^2 &=  M \frac{\sigma_i^2}{\triangle_i^{'}},
\end{align}
\end{subequations} 
with
\begin{equation}
\label{eq:triangle}
\triangle_i^{'} =  M \sum_{m=1}^M x_{i,m}^2-\left(\sum_{m=1}^M x_{i,m} \right)^2.
\end{equation} 
The prime superscript indicates that the variance of the measurements 
$\sigma^2_{i,m}=\sigma^2_{i}$ for the $i$-th MCC was assumed to be constant.
It was estimated as the variance $s^2$ of the sample population:
 \begin{equation}
\label{eq:sample-variance}
\sigma_{i}\approx s^2=\frac{1}{N-2}\sum_{m=1}^M (y_m-a_i-b_i x_m)^2.
\end{equation}
We normally imaged $M=4$ conditions per MCC. For exceptions with $M=2$, $s^2$ could not be computed directly from (\ref{eq:sample-variance}) and was assumed to equal the largest value from the other experiments.

\begin{figure*}
\begin{center}
\includegraphics[width=0.8\linewidth]{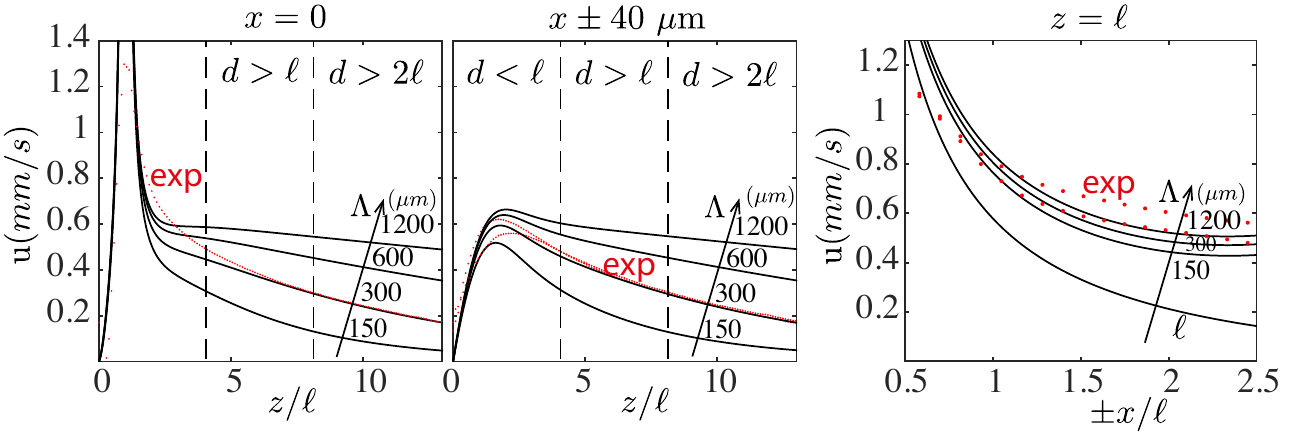}
\caption{The lateral velocity of a two-dimensional array of Stokeslets of radius $\Lambda \sim 300$, falls off as in experiments (exp), as shown for:  (a,b) above and between bundles, respectively, as a function of $z$; (c) between bundles as a function of $x$.  Results are obtained by direct summation of the exact solution of each Stokeslet, all of strength $F_c\mathbf{e}_x$ and z-offset $\ell$.}
\label{fig:Sfig3}
\end{center}
\end{figure*}

\begin{figure}[t]
    \centering
    \includegraphics[width=1.0\linewidth]{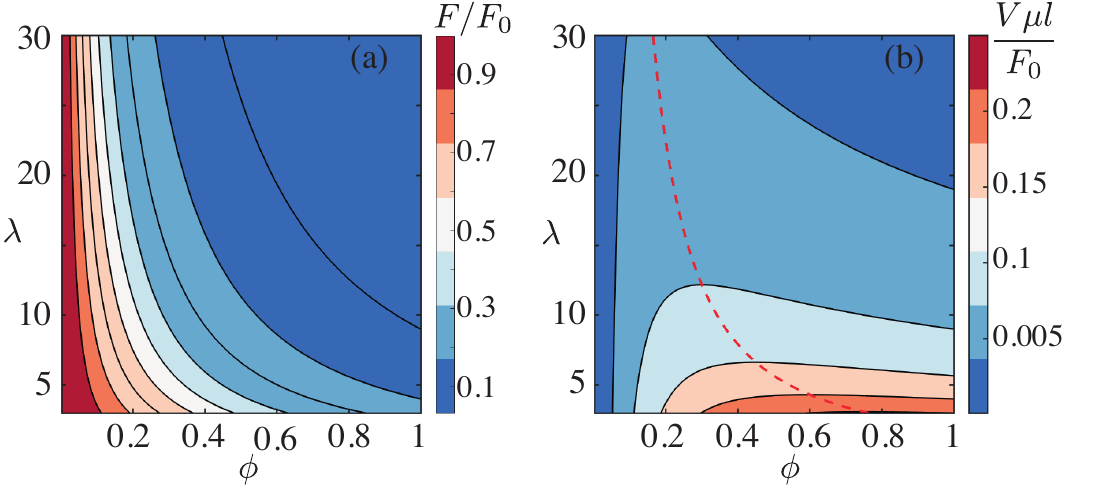}
    \caption{Supplement of Fig.\ 4. Contour plot of (a) the effective force $F$ [Eq.\ (5)], and (b) the limit velocity $V$, in parameter space.  Dashed red line in (b) traces ridge of optimization for $V$.}
    \label{fig:Sfig4}
\end{figure}

\medskip

{\bf Fitting the near flow-field by the singularity method:}
The flow $u_c$ driven by the cilia in $\Omega_c$  is modelled as the superposition of the flows arising from 
local point forces (Stokeslets) $\mathbf{f}_n$ applied at $\mathbf{s}_n \in \Omega_c$:   
\begin{equation}
\label{eq:singMethod}
\mathbf{u}_c(\mathbf{b}_i) \approx \sum_{n=1}^N \mathbf{f}_n \mathbf{\cdot} \mathbf{S}(\mathbf{b}_i-\mathbf{s}_n). 
\end {equation}
The tensor   $\mathbf{S}$  is the well-known, exact solution for a Stokelset next to a no-slip plane 
at $z=0$ \cite{Blake1971SM}. 
The values of  $\mathbf{f}_n$ are found by fitting $\mathbf{u}_c$ at $M$ collocation points $\mathbf{b}_i$, with $M>2N$ 
to avoid numerical instabilities \cite{Boselli2012a}. As no-slip boundary conditions $\mathbf{u}_c=0$ at $z=0$ are 
implicitly satisfied, walls do not need to be discretized. 
The linear system \eqref{eq:singMethod} is then simply recast in its matrix form $\mathbf{Af} = \mathbf{u}_b$, with the $3M\times 2N$ matrix
$$\mathbf{A}=\left(\begin{array}{cc}S_{11}(\mathbf{b}_i,\mathbf{s}_j) & S_{13}(\mathbf{b}_i,\mathbf{s}_j) \\S_{21}(\mathbf{b}_i,\mathbf{s}_j) & S_{23}(\mathbf{b}_i,\mathbf{s}_j) \\S_{31}(\mathbf{b}_i,\mathbf{s}_j) & S_{33}(\mathbf{b}_i,\mathbf{s}_j)\end{array}\right)$$
the $2N\times 1$ vector 
$\mathbf{f} = \{f_{1,1}, \ldots, f_{N,1}, f_{1,3}, \ldots, f_{N,3} \},$ 
and the $3M\times 1$ vector  $\mathbf{u}_b=\{u_{c,1}(\mathbf{b}_i), u_{c,2}(\mathbf{b}_i), u_{c,3}(\mathbf{b}_i) \}.$  
We then solve for $\mathbf{f}$ using the backslash operator of Matlab.

We set $f_2 = 0$, assuming the solution to be symmetric in $y$. 
Once $\mathbf{f}_n$ are known, \eqref{eq:singMethod} can be used to evaluate the fitted solution at any $\mathbf{x}$. 
The flow field measured in the $y$ plane is extruded by replications at $13$ planes evenly spaced between 
$-10\,\mu{\rm m} < y < 10\,\mu$m. We used $15$ Stokeslets for each plane about the fictitious boundary 
$\partial \Omega_c$.  

\medskip

{\bf Coarse-graining the bundle:}
The flow $\mathbf{u}_c$ driven by the Stokeslet in the bundle can be coarse-grained further, 
with a smaller number of Stokes flow singularities, moving away from the bundle.  
We compare the fitted flow $\mathbf{u}_c$, made up of $N$ Stokeslets as discussed above, with the flow driven by the 
effective Stokeslet $F_c\mathbf{e}_x$ applied at $(0,0,\ell)$, and the effective  rotlet $2\ell F_c \mathbf{e}_y$ applied 
at $(0,0,\ell/2)$. 
They share the same far-field, reflecting the fact it is driven by an active vortex.  The flow driven by the 
entire bundle decays as $1/z^3$, as for a single singularity, for $z>2\ell$ (Fig.\ \ref{fig:Sfig1}).

\medskip

{\bf Far Field fitting:}
The flow given by (2),
is used to fit the PIV measurements for $z>2d$ (Fig.\ \ref{fig:Sfig2}).
The velocity $u_{far}(z;\Lambda)$ depends linearly on $V$, but not on $R$. For a given value of $R$, we find 
$V$ by a linear least-squares fit of the data. We then simply repeat this linear fit for candidate values 
in the range $30\,\mu{\rm m} < R < 1\,$mm, with increment $\Delta R = 10\,\mu$m, and select the value of 
$R$ that minimizes the $L_2$ fitting error. 

\medskip

{\bf Two-dimensional array of Stokeslets:}
Results similar to those presented in Fig.\ 2 
for a uniform distribution of Stokelets can be obtained by positioning Stokeslets $F_c\mathbf{e}_x$ on a 
lattice with cut-off radius $\Lambda$. Each element $\mathbf{s}_{ij}$ of the lattice is position 
at $(x_{ij} = id_{11}  + j d_{12}, y_{ij} = j d_{22}, z_{ij}= \ell)$.  
From confocal imaging of the closest neighboring cells of the bundle in Fig. 2,
we estimate  $d_{11} \sim 70\,\mu$m, $d_{12} = 40.5\,\mu$m, $d_{22} = 53 \mu m$. 

Using the effective force $F_c$ estimated by the near field fitting, and summing up the exact 
contribution of each MCC, we retrieve the slow decay rate observed {\it in vivo} [Figs. 2(e,f)] 
for $\Lambda \sim 300\,\mu$m (Fig.\ \ref{fig:Sfig3}). This is the same result found by fitting the 
far-field flow with Eq. (2).

\medskip

{\bf Resistive force theory estimate of the effective force applied by a single cilium:}
We adopt a simplified view of the power stroke of a cilium as a straight rod that pivots around its base. 
Let $s\in [0,\ell]$ be arclength along a cilium, with $s=0$ at the base and $s=\ell$ at the tip, 
and let $\phi$ be the angle between the cilium and the wall.  
The lateral component of the RFT force density at  $s$  is $f' \sim (s/\ell) \zeta_{\sst\perp}V_c \sin{\phi} $, and the 
resulting far field velocity,  given by Eq. (1), is proportional to $h f' ds$ with $h=s\sin{\phi}$.
Accordingly, the effective force $f_{\phi} = \ell^{-1} \int_0^\ell  h(s,\phi) f'(s,\phi)  ds $, 
matches the overall far field 
$\ell f_{\phi} \tilde{S}_{11}$ when applied at $\ell$. We obtain 
$f_{\phi}=  \sin^2{\phi}\,\, \zeta_{\sst\perp} \ell V_c/3$.  Through the entire stroke, 
a cilium cycles through an angle $\Delta \phi = 2\pi$, and we assume that the recovery stroke does not
contribute to the force, so $f_\phi=0$ for $\pi<\theta<2\pi$. Averaging  $f_{\phi}$ gives the effective force 
$f = \zeta_{\sst\perp} \ell V_c (6\pi)^{-1}\int_0^\pi \sin^2{\phi} \, d\phi$, which gives the expression $f=\zeta_{\sst\perp} \ell V_c /12$ used in the main text. 
\medskip

{\bf Additional results from the self-consistent model:}
The self-consistent model in Eq. (5) can be used to investigate several aspects of the phenomenology of 
cilia driven flow.  
The contour plots for the effective force $F$ and the limit velocity $V$ are shown in Fig.\ \ref{fig:Sfig4}.
$V$ has a maximum at $d = (\lambda/2)^{1/3}\ell\approx 32\, \mu$m. The corresponding coverage fraction 
$\phi = (\lambda/2)^{-2/3} \sim 0.22$ is significantly larger than observed \textit{in vivo}, confirming 
that the system is instead optimized for the wall force $F_w$ (Fig. 4) discussed in the main text. 

\medskip

{\bf Supplementary video:}
Movie of a cilia bundle and $0.2\,\mu$m diameter tracers, acquired at $2,000$ frames/s, and shown 
at $30$ frames/s. Some larger beads are also present to help visualize the flows.


%

%

\end{document}